\documentclass[aps,prl,twocolumn,showpacs]{revtex4}
\usepackage{graphicx}
\usepackage{dcolumn}
\usepackage{bm}

\begin{document}

\title{Unification of bulk and interface electroresistive switching in oxide systems}

\author{A. Ruotolo}
\author{C. W. Leung}
\author{C. Y. Lam}
\author{W. F. Cheng}
\author{K. H. Wong}

\affiliation{Department of Applied Physics and Materials Research Centre, The Hong Kong Polytechnic University, Hung Hom, Kowloon, Hong Kong, China}

\author{G. P. Pepe}
\affiliation{CNR-INFM Coherentia and Dipartimento Scienze Fisiche, Universit\`a di Napoli "Federico II", Piazzale Tecchio 80, 80125 Naples, Italy}

\date{\today}

\begin{abstract}
We demonstrate that the physical mechanism behind electroresistive switching in oxide Schottky systems is electroformation, as in insulating oxides. Negative resistance shown by the hysteretic current-voltage curves proves that impact ionization is at the origin of the switching. Analyses of the capacitance-voltage and conductance-voltage curves through a simple model show that an atomic rearrangement is involved in the process. Switching in these systems is a bulk effect, not strictly confined at the interface but at the charge space region.
\end{abstract}


\maketitle

Reversible electroresistive (ER) switching effect in oxide Schottky junctions has been widely reported in the last years  \cite {Zhang:APL:2005,Kim:JJAP:2005,Fujii:PRB:2007}. In the effort to claim potentiality for non-volatile memory applications, it has been sustained that, unlike ER switching in insulating oxides \cite{Dearnaley:RPP:1970}, the effect is electronic and hence very fast. As recently pointed out by Waser and Masakazu Aono  \cite{Waser:NatMater:2007}, unambiguously experimental evidences have not been provided in this direction. Moreover, the papers often lack of details on fabrication processes and measurement setups. As a consequence, it is difficult to draw boundaries for possible applications.

The proposed scenario is charging effect at the Schottky interface \cite {Fujii:PRB:2007} through Fowler-Nordheim tunneling to defect states: acceptor trapping states are present in the Shottky barrier because of impurities and structural defects. In the high resistance state these traps are occupied with electrons. When a sufficiently high forward bias voltage is applied, electrons are detrapped, leaving resonant tunneling paths in the Schottky barrier; as a consequence, the resistance drops. A negative reverse bias voltage is able to produce retrapping, hence to switch back the device to the previous state.
In other words, resonantly tunneling paths into the Schottky barrier would play the same role as filamentary percolative paths created by electroformation in insulating oxides \cite{Szot:NatureMat:2006}.

In this report we give experimental evidences sustaining electroformation at the origin of the effect, as in bulk undoped oxides  \cite{Waser:NatMater:2007,Szot:NatureMat:2006}. The trapping states are not due to impurities or structural defects but created by avalanche injection breakdown.  Ionic motion is involved, although the atomic rearrangement remains confined below the interface over a length scale of the space charge.

The system La$_{0.7}$Sr$_{0.3}$MnO$_{3}$/SrTi$_{0.98}$Nb$_{0.02}$O$_{3}$ (LSMO/NSTO) was chosen for the present investigation. A full characterization of the interface transport properties of this system has been already reported \cite{Ruotolo:PRB:2007}. To avoid any possible additional capacitance contribution from patterned wiring when measuring at high frequency, for the present study we chose a capacitor-like geometry. A 100 nm thick LSMO film was deposited on a 0.5 mm thick NSTO substrate through a $500 \times{500}~\mu{}m^{2}$ shadow mask and, subsequently, a Platinum film was evaporated on the LSMO for the electrical contact. In the following, the positive voltage bias is defined by the current flowing from LSMO to NSTO.

\begin{figure}[t]
\includegraphics[width=\columnwidth]{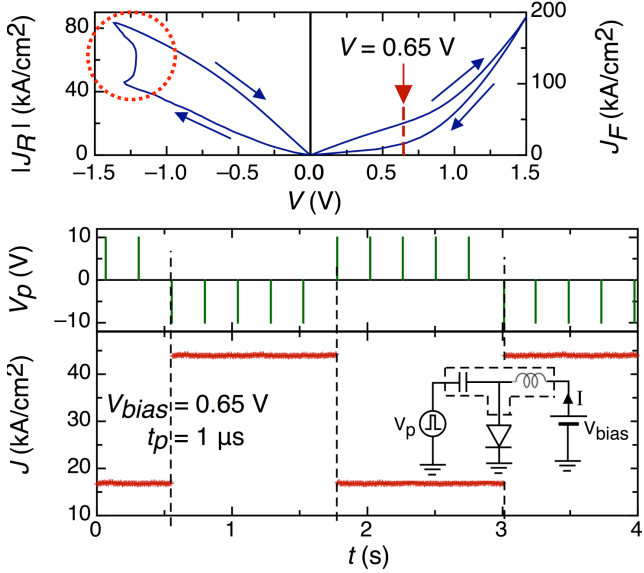}
\caption{\label{fig1}(Color online) Top: Hysteretic current density - voltage ($J-V$) characteristic. The forward bias current ($J_{F}$) and the absolute value of the reverse bias current ($J_{R}$) are reported on two different scales. The $J-V$ shows negative resistance when reverse switching occurs. Bottom: ER switching induced by voltage pulses with a duration of $t_{p}=1~\mu{}$s. The ER ratio measured in the pulsed working mode is the same as that evaluated from the  $J-V$, at the same bias. Inset shows the measurement setup.}
\end{figure}

The junctions could be switched between two stable states (Fig.~\ref{fig1}).
Switching towards the low resistance state (LRS) can be achieved by either scanning the negative voltage (reverse bias) or by applying a short negative voltage pulse. The application of a subsequent train of pulses of the same sign does not change the state of the device. Switching back towards the high resistance state (HRS) can be achieved by either scanning the positive voltage (forward bias) or by applying a positive voltage pulse. Like for the LRS, the HRS is stable over the application of further positive pulses. In the pulsed working mode, the ER switching was detected by reading the current at a fixed voltage. A standard bias tee was used to supply simultaneously the pulses and the bias voltage to the junction. The dc current was measured as the current flowing through the dc power supplier. The ER ratio was the same as that evaluated from the hysteretic current-voltage ($I-V$) curve, at the same bias. The dipolar characteristic of the effect, together with the large compliance current of the voltage suppliers ($\sim$ 100  mA), rules out Joule heating as the dominant mechanism  \cite{Waser:NatMater:2007}.

\begin{figure}[t]
\includegraphics[width=\columnwidth]{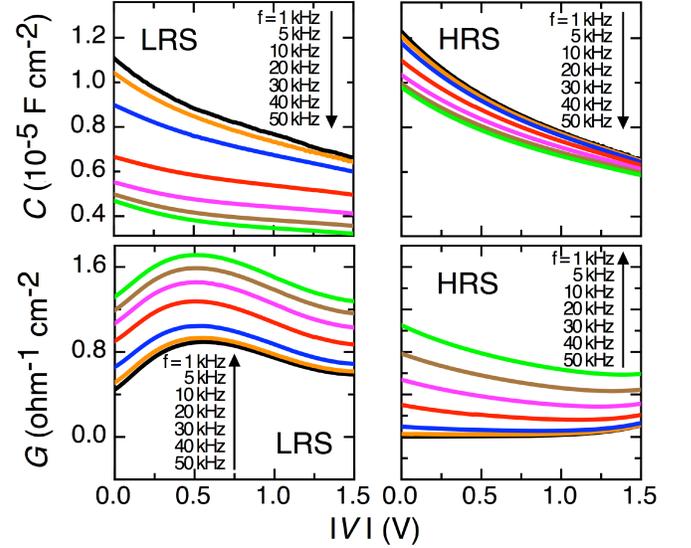}
\caption{\label{fig2}(Color online) $C-V$ and $G-V$ curves in the two resistance states recorded in reverse bias at different values of $f$: 1, 5, 10, 20, 30, 40 and 50 kHz from top to bottom in the $C-V$'s and bottom to top in the $G-V$'s.}
\end{figure}

The presence of trapping states at the interface of a Schottky junction can be detected by measuring the capacitance ($C$) and the conductance ($G$) as a function of $V$ with frequency ($f$) as parameter \cite{Sze:Physics_of_semiconductor_devices}.  Fig.~\ref{fig2}
shows the $C-V$'s and the $G-V$'s taken at various $f$ in reverse bias (superimposed oscillating voltage of 25 mV) and in both the resistance states. Such a combination of $C-V$'s and $G-V$'s is peculiar in the framework of standard semiconducting Schottky junctions. A shift of the $C-V$'s with $f$ is typically detected in metal-insulating oxide-semiconductor (MIS) systems and can have several origins. The most common is the presence of interface states at the I - S interface, where the periodic lattice structure is interrupted \cite{Nicollian:APL:1965}. We do not have an insulating oxide in our system. Yet, the semiconductor is itself an oxide, therefore one could imagine that dangling Ti bonds at the interface act as acceptor states. As a matter of fact, when increasing $f$, the $C-V$ shifts towards less negative (more positive) voltage, as expected for acceptor states at the interface \cite{Sze:Physics_of_semiconductor_devices}. In the LRS, the overlapping between Ti acceptor states results in the formation of a conducting band in the mid-gap. When a voltage is applied, the peak in the $G-V$ curve, corresponds to the Fermi level crossing the interface state band in the gap. 

This simple scenario is ruled out by the weak shift of the peak position with $f$ in the $G-V$'s. In case of interface trapped charge, the position of the peak in the $G-V$ is expected to shift accordingly with the shift of the $C-V$ because both are function of surface (interface) potential. In Fig.~\ref{fig2} (bottom-left) the peak position is weekly sensitive to the frequency. If trapping states exist they are not strictly confined at the interface. 

Another possibility is the presence of trapping centers in the depletion layer due to impurities or crystal defects. This can also be excluded with a simple calculation. When the ac voltage swings, at high frequency trapped charges cannot follow it, thus, when $f$ increases, the $C-V$ approaches the ideal curve, free of capacitance due to traps. Choosing a fixed value of $C=6.6\times 10^{-6}$ F/cm$^{2}$, a change of $f$ from 1 kHz to 20 kHz produces a shift of about 1.5 V in the LRS. This means that the trapping density per unit area $Q_{tot}$ is larger than $\Delta Q =  (6.6\times 10^{-6}) \times 1.5 = 10^{-5}$ C/cm$^{2} = 6.25 \times 10^{13}$ charges/cm$^{2}$. 
The built-in depletion width $W_{D0}=W_{D}(V =0)$ of a Schottky junction can be calculated from \cite{Sze:Physics_of_semiconductor_devices}: 
\begin{equation}
W_{D} = \sqrt {\frac {2 \epsilon_{S} (V_{B0}-V)} {qN_{S}}} 
\label{eq1}
\end{equation}
where $q$ is the electron charge,  $V_{B0}$ is the built-in potential, $N_{S}$ is the doping concentration and $\epsilon_{S}$ is the permittivity in the depletion layer. In our as-grown junctions \cite{Ruotolo:PRB:2007}, where $N_{S}=3.3 \times 10^{20}$ cm$^{-3}$, $V_{B0}=0.83 \pm 0.05$ V and $\epsilon_{S}=54.4$ at room temperature, $W_{D0}=3.9$ nm. This implies that a concentration of defects larger than $N_{d}=\Delta Q/W_{D0} = 1.6 \times 10^{20}$ cm$^{-3}$ should exist in the original crystal, which is absolutely unlikely. Notice how $N_{d}$ is suspiciously very close to $N_{S}$.
If there were no ovelapping between defect states (fixed charges), the shift of the $C-V$ from LRS to HRS should be rigid and no peak should be detected in the $G-V$ in either states.

We can simply conclude that the ER effect is not due to any kind of structural defects of the original semiconducting crystal, neither at the interface nor in the depletion layer. It must be the application of a voltage above a certain threshold that creates trapping states and possibly fixed charge in the crystal. 

In Fig.~\ref{fig1}, when the reverse threshold voltage ($V_{sw,r}$) is reached, the $I-V$ shows S-type negative differential resistance \footnote{A resistor is used in series as a protection against forward breakdown when sweeping the $I-V$, therefore the diode is actually biased in current during this phase.}. Only after that a negative resistance is recorded, the $I-V$ becomes hysteretic. The appearance of an S-type negative differential resistance is a common observation in Schottky contacts and it is due to avalanche injection under the application of a large negative electric field \cite{Gunn:ProcPhysSocB:1956}. In standard semiconductors, avalanche ionization begins when the electric field reaches values of the order of $10^{5}$ V cm$^{-1}$ \cite{McKay:PR:1953}, \emph{i.e.} large voltage bias are required. In our junctions, the built-in electric field at the interface is already $E_{max}= 2(V_{B0}/W_{D0})= 4.1 \times 10^{6}$ V cm$^{-1} $. As a consequence, the avalanche ionization has a very low threshold voltage. In MIS diodes, avalanche injection causes a flat-band shift in the $C-V$ towards more positive voltage, the appearance of a peak in the $G-V$ and a reversible change in the $I-V$ \cite{Sze:Physics_of_semiconductor_devices}. The shift indicates an increase of negative trapped charge in the insulating oxide and the appearance of the peak indicates that the charge is mainly trapped close to the I - S interface, where the plasma is confined. Some of the electrons have enough energy to surmount the I - S interfacial energy barrier and enter the insulating oxide where they remain trapped and form a midgap band. In our system, we do not have an insulating oxide where electrons can be trapped. Therefore, the negative charge or acceptor states must be created by impact ionization. This can happen in only one way: break of bondings and atomic rearrangement. 

\begin{figure}[t]
\includegraphics[width=\columnwidth]{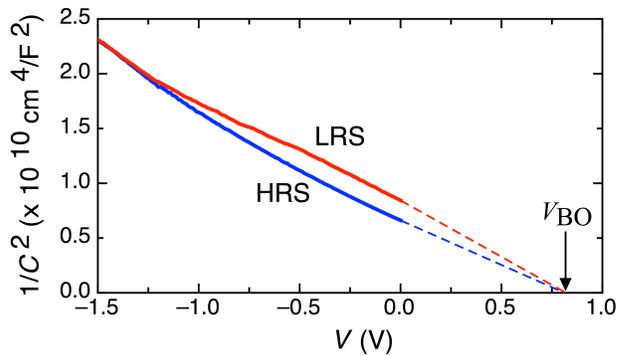}
\caption{\label{fig3} (Color online) $1/C^{2}-V$ curves at $f=1$ kHz. The built-in potential is the same in both states.}
\end{figure}

When a negative voltage is applied, the electric field close to the interface reaches a value at which carriers have enough energy to ionize some of the valence bonds of the lattice and to liberate one, or more than one, mobile electron-hole pairs. The new carriers can cause further ionization that would lead to irreversible breakdown if any feedback mechanism intervened to switch off the plasma. The avalanche ionization must be stopped because of a sudden drop of the electric field. A drop of electric field can occur because of a change in the barrier height and/or in the barrier width. 
If we plot (Fig.~\ref{fig3}), 
$1/C^{2}-V$ at low frequency in the LRS and HRS, the curves are linear at low reverse bias voltage and $V_{B0}$ can be determined as the intercept with the $V$ axis \cite{Sze:Physics_of_semiconductor_devices}:
\begin{equation}
\frac {1}{C^{2}} =  2 \frac { V_{B0} - V} {q \epsilon_{S}  N_{S}}   
\label{eq2}
\end{equation}

$V_{B0}$ is the same in the two states (and equals to that measured in as-grown junctions).
When increasing the reverse bias voltage the curves deviate from the linear behavior. For the HRS this deviation is of the same type of that commonly recorded in as-grown all-oxide Schottky junctions \cite{Ruotolo:PRB:2007,Sawa:APL:2005} and can be attributed to reduced measurement accuracy due to the increase of the reverse leakage current. Instead, the behavior in case of LRS is peculiar and indicative of an anomalous non-uniform distribution of space charge. The curve in this case goes through an S-type bending to join the HRS curve at a voltage that corresponds to $V_{sw,r}$. This suggests that the semiconductor is affected by the switching only over a distance from the interface $W_{sw,r}=W(V_{sw,r}=$ -1.3 V$)=$ 6.3 nm, as calculated from Eq.~\ref{eq1}.

Impact ionization produces negative charge in the depletion layer. If the built-in potential is not affected, the presence of extra-negative charge is compensated by spontaneous ionization of a larger number of donors, \emph{i.e.} an increase of $W_{D}$. An increase of $W_{D}$ implies a reduction of $C$, consistently with the $C-V$ shift observed when switching from HRS to LRS. When $W_{D}$ increases, the electric field decreases until the plasma switches off preventing breakdown. In these systems, avalanche injection is self-feedbacked.

Notice that in the LRS  $W_{D,LRS}$ can not be calculated by evaluating the permittivity from the slope of $1/C^{2}-V$ at $V = 0$ and applying Eq.~\ref{eq1} as is. The slope is larger at $V=0$ but much smaller at higher reverse bias voltage, as compared to that in the HRS. Instead, in the HRS the permittivity and the built-in depletion width were evaluated to be $\epsilon_{S,HRS}=54.1$ and  $W_{D,HRS}=3.9$ nm. 

$W_{D,LRS}$ must be larger than $W_{D,HRS}$ but of the same order of magnitude, if the semiconductor is affected only over a distance $W_{sw,r}=6.3$ nm. On the other hand, since all the negative charge created during ionization must be compensated by spontaneous ionization of Nb, $W_{D,LRS}$ being of the same order of $W_{D,HRS}$ is consistent with the previously calculated trap density $N_{d}$ being of the same order of the doping level $N_{S}$.

It remains to understand which bondings are broken during ionization. Any uniform change of stoichiometry into the compound can only produce a rigid shift of the Fermi energy. A conducting band can appear in the midgap in only one case: if clustering of oxygen vacancies occurs \cite{Shanthi:PRB:1998}. The scenario must be the following. Ti - O bonds are broken during ionization. The negative electric field exerts its force on the negative O-ions in such a way to displace them far from the interface (not further than $W(V_{sw,r})$) and leave Ti acceptor states close to the interface. When the external reverse voltage is brought back to zero, the (still) negative built-in electric field acts against the oxygen gradient force, hence inhibiting the diffusion of O-ions in the opposite direction. In these state, overlapping between neighbor Ti acceptor states results in a conducting band in the mid-gap. The $G-V$ detects only the surface states \cite{Nicollian:APL:1965}, with the result that the shift in the $G-V$ with $f$ is small while the large shift of the $C-V$ reveals the presence of a large concentration of oxygen ions deep into the depletion layer. 

In order to switch the device back, a positive voltage must be applied to compensate (at least) the negative internal electric field. As a matter of fact, a voltage larger than $V_{B0}$ (see Fig.~\ref{fig1}) must be applied to induce hysteresis in the $I-V$. Compensation gives to the oxygen the possibility to slowly diffuse back. This is consistent with the absence of a well-defined forward threshold voltage when slowly sweeping the $I-V$. The switching is bursted up by applying fast voltage pulses larger than $V_{B0}$. In our system the application of a voltage larger than $V_{B0}$ (that would produce irreversible breakdown in a standard diode) is possible because of the presence of the conducting paths formed along the oxygen dislocations. The space charge region can be seen as a three-dimensional network of diodes and metallic resistors \cite{Benguigui:PRB:1988,Szot:NatureMat:2006}.

Notice that, the device never goes back to the as-grown state. If the device switched back to the as-grown state, suddenly a voltage much larger than the barrier height would be applied to an as-grown diode, which would completely destroy it. From a microscopical point of view this means that, when the device is switched back to the HRS, the previously formed oxygen vacancies are not completely refilled but conducting paths are all interrupted, exactly in the same way as it occurs in filamentary type switching in undoped STO \cite{Szot:NatureMat:2006}. Dislocations are still present in the depletion layer although conduction is inhibited. This explains why in Fig.~\ref{fig2} a shift of the $C-V$ with $f$ is still observed  in the HRS, but no peaks are detected in the corresponding $G-V$'s.  

We can now draw some boundaries on the size and the time-scale limitations of these devices for non-volatile memory applications. The semiconductor is affected by the switching only over a distance below the interface $W_{sw,r}$. As an upper limit, we can assume that the plasma does not extend outside the interface area further than $W_{sw,r}$ along the interface plane \cite{Sze:Physics_of_semiconductor_devices}. For doping level of the order of that used in this work, the minimum distance between two nearest neighbor bit cells can be of the order of a few nm. Large values of $N_{S}$ imply small values of depletion width and large built-in electric fields. As a consequence, the bit density increases and, at the same time, the switching threshold decreases. This explains the large increase of reverse threshold voltage reported in Ref.~\cite{Fujii:PRB:2007} when reducing $N_{S}$. On the other hand, the ER ratio decreases with $N_{S}$  \cite{Fujii:PRB:2007}. This can be explained with a smaller built-in electric field, and hence a smaller gradient of charge distribution in the depletion layer, \emph{i.e.} a smaller number of dislocations.  

As far as the switching time is concerned, determining the mobility of the O-ions along the conducting paths is mandatory. We can only observe here that the avalanche multiplication is a very fast process ($\sim 10$ ps in standard semiconductors \cite{McKay:PR:1953}) and that, once the bondings have been broken, O-ions are accelerated by a very large electric field to be displaced over lenght-scale of the depletion width,  \emph{i.e.} a few nm. It is not unreasonable to believe that these devices can reach, or even overcome, the working frequency of the current non-volatile memories.

In conclusion, we have demonstrated that the ER switching in Shottky contacts is due to electroformation, as in bulk insulating oxides. Yet, the high electric field at the Schottky interface confines the plasma, and hence the effect, to a specific volume below the junction area, opening up the path to high density integration of resistive random access memories. Moreover, the presence of a built-in electric field reduces the switching voltage and makes the LRS stable over a very long retaining time.

This work was supported by the Hong Kong Polytechnic University through Grants No. A-PG91 and 1-ZV43. We acknowledge A. Cassinese and coworkers from the University of Naples for discussions.

\end{document}